\documentstyle[twocolumn,aps,epsf,floats]{revtex}

\newcommand{\bi}[1]{\mbox{\boldmath$#1$}}

\begin{document}

\draft

\title{Anomalous spatio-temporal chaos in a two-dimensional system of\\ 
  non-locally coupled oscillators}

\author{Hiroya Nakao\footnote{E-mail:nakao@ton.scphys.kyoto-u.ac.jp}}

\address{Department of Physics, Graduate School for Sciences, Kyoto
  University, Kyoto 606-8502, Japan}

\date{March 8, 1999}

\maketitle

\begin{abstract}
  A two-dimensional system of non-locally coupled complex
  Ginzburg-Landau oscillators is investigated numerically for the
  first time.
  As already known for the one-dimensional case, the system exhibits
  anomalous spatio-temporal chaos characterized by power-law spatial
  correlations.
  In this chaotic regime, the amplitude difference between neighboring
  elements shows temporal noisy on-off intermittency.
  The system is also spatially intermittent in this regime, which is
  revealed by multi-scaling analysis; the amplitude field is
  multi-affine and the difference field is multi-fractal.
  Correspondingly, the probability distribution function of the
  measure defined for each field is strongly non-Gaussian, showing
  scale-dependent deviations in the tails due to intermittency.
\end{abstract}



{\bf Assemblies of mutually interacting dynamical units are ubiquitous
  in nature. As models for them, systems of coupled simple dynamical
  elements, e.g. limit cycle oscillators or chaotic maps, have been
  studied extensively.
  Recently, a new class of coupled systems, namely, a system with
  non-locally coupled dynamical elements, was introduced and found to
  exhibit anomalous spatio-temporal chaos characterized by power-law
  spatial correlations.
  The preceding studies on this system have been done in one
  dimension, but the mechanism for the appearance of this
  spatio-temporal chaos seems to be universal, and it is also expected
  in higher dimensions.
  In this paper, a two-dimensional system of non-locally coupled
  oscillators is investigated for the first time. As in the
  one-dimensional case, the system is found to exhibit anomalous
  spatio-temporal chaos, accompanied by several distinctive features
  specific to this chaotic regime, i.e., power-law spatial
  correlations, noisy on-off intermittency, and multi-scaling
  properties.  }

\section{Introduction}

Assemblies of dynamical elements coupled with each other are widely
seen in nature. Simplified models of such systems, e.g. coupled limit
cycle oscillators or chaotic maps, have played important roles not
only in modeling such systems realistically, but also in understanding
the varieties of possible behavior of systems far from
equilibrium. Many important concepts, such as pattern formation or
spatio-temporal chaos, have been extracted from the detailed studies
on such models.

The interaction between the elements are usually assumed to be
attractive, and of mean-field type in a wide sense; each element feels
the mean amplitude of its neighboring elements, and driven by the
difference between its amplitude and the mean amplitude, in such a way
that the amplitude differences between the elements decrease.

The diffusive coupling is a representative limiting case. Each element
interacts strongly with its nearest neighbors, so that the amplitude
field of the system is always continuous and smooth. It is well known
that some diffusively coupled systems of dynamical elements, such as
the complex Ginzburg-Landau equation, exhibit spatio-temporal
chaos\cite{Kuramoto2,Bohr}.

The opposite limiting case is the global coupling, or the mean field
coupling in the narrow sense. Each element feels the mean field of the
entire system, and is thus coupled to all the elements with equal
strength. The amplitude field becomes statistically spatially
homogeneous, and the notion of space is lost. It is known that systems
with global coupling generally show some typical behavior,
e.g. clustering and collective chaos\cite{Hakim,Kaneko}.

In Ref.\cite{Kuramoto1}, Kuramoto introduced an intermediate system
between the above two limiting cases, namely a system of non-locally
coupled elements. Our subsequent numerical simulations of
one-dimensional non-locally coupled systems with various elements
revealed that such systems generally exhibit anomalous spatio-temporal
chaotic behavior, which cannot be seen in the above two limiting
cases.
In this chaotic regime, the amplitude field becomes fractal, and the
spatial correlation of the amplitude field shows power-law behavior on
small scales. Furthermore, the fractal dimension and the exponent of
the spatial correlation vary continuously with the coupling strength.
Later, we developed a theory that can explain the fractality of the
amplitude field and the power-law behavior of the spatial correlation
based on a simple multiplicative stochastic
model\cite{Kuramoto3,Nakao1}. Such a model is frequently employed in
describing the noisy on-off intermittency
phenomena\cite{Pikovsky,Platt,Cenys} found in many physical systems,
and this implies that our system should also exhibit this type of
temporal intermittency.
Induced by the temporal intermittency, our system is also spatially
intermittent. In order to study this, we expanded our analysis of the
amplitude field into more general $q$-th structure functions, and
found that the amplitude field shows multi-affinity. We also
introduced multi-fractal analysis of the difference field of the
original amplitude field, inspired by its seeming similarity to the
intermittent energy dissipation field in fluid
turbulence\cite{Kuramoto3,Nakao2}.

All our previous studies have been done in one-dimensional systems.
However, our previous theory does not require the systems to be
one-dimensional, and spatio-temporal chaos with power-law structure
functions is also expected in higher dimensions.
In this paper, we study a system of non-locally coupled complex
Ginzburg-Landau oscillators in two dimensions for the first time, and
investigate its anomalous spatio-temporal chaotic regime. Some
attention is paid to the multi-scaling properties of the intermittent
amplitude and difference fields.

\section{Model}

As proposed by Kuramoto\cite{Kuramoto1}, the non-local coupling
naturally appears in the following plausible situation.
Consider an assembly of spatially distributed dynamical elements, e.g.
cells. Each element is assumed to interact indirectly with other
elements through some (e.g. chemical) substance, which diffuses and
decays much faster than the dynamics of the element.
Such a situation will be described by the following set of equations:
\begin{eqnarray}
  \dot{\bi X}({\bi r}, t) &=& {\bi F}({\bi X}({\bi r}, t)) + {\bf K}
  \cdot {\bi A}({\bi r}, t),
  \label{Eq:01} \\ 
  \epsilon \dot{\bi A}({\bi r}, t) &=& -\eta {\bi A}({\bi r}, t) + D
  \nabla^2 {\bi A}({\bi r}, t) + {\bi X}({\bi r}, t),
  \label{Eq:02}
\end{eqnarray}
where ${\bi X}$ is the amplitude of the element, ${\bi F}$ is the
dynamics of the amplitude, and ${\bi A}$ is the concentration of the
substance with decay rate $\eta$ and diffusion rate $D$.
The substance ${\bi A}$ is generated at a rate proportional to the
amplitude ${\bi X}$, and the amplitude ${\bi X}$ is affected by the
substance ${\bi A}$ with a coupling matrix ${\bf K}$.
The parameter $\epsilon$ determines the ratio of time scale of the
elements to that of the substance, and is assumed to be very
small. Namely, the dynamics of the substance is much faster than that
of the elements.

Now, let us consider the $\epsilon \to 0$ limit and eliminate the
dynamics of ${\bi A}$ adiabatically. Putting the left-hand side of
Eq.(\ref{Eq:02}) to $0$, we can solve the equation for ${\bi A}$ as
\begin{equation}
  {\bi A}({\bi r}, t) = \int d{\bi r}' G({\bi r}' - {\bi r}) {\bi
    X}({\bi r}', t),
  \label{Eq:A}
\end{equation}
where $G({\bi r}'-{\bi r})$ is a kernel that satisfies
\begin{equation}
  (\eta - D \nabla^2) G({\bi r}' - {\bi r}) = \delta( {\bi r'} ).
\end{equation}
By inserting Eq.(\ref{Eq:A}) into Eq.(\ref{Eq:01}), we obtain the
following system of non-locally coupled dynamical elements:
\begin{equation}
  \dot{\bi X}({\bi r}, t) = {\bi F}({\bi X}({\bi r}, t)) + {\bf K}
  \cdot \int d{\bi r}' G({\bi r}' - {\bi r}) {\bi X}({\bi r}', t).
\end{equation}

The kernel $G({\bi r}'-{\bi r})$ can be solved as
\begin{equation}
  G({\bi r}' - {\bi r}) = \frac{1}{(2 \pi)^d} \int d^d {\bi q} 
  \frac{\exp[i {\bi q} \cdot ({\bi r}' - {\bi r})]}{\eta + D |{\bi q}|^2},
\end{equation}
where $d$ is the dimension of the space.
When the system is isotropic, the kernel $G$ becomes a function of the
distance $r : = |{\bi r}' - {\bi r}|$, and is expressed as
\begin{eqnarray}
  G(r) \propto& \exp( -\gamma |r| ) &\;\;\;\; \mbox{d=1}, \label{Eq:Kernel1d}\\
  & K_0( \gamma |r| )   &\;\;\;\; \mbox{d=2}, \label{Eq:Kernel2d} \\
  & \displaystyle{\frac{\exp( -\gamma |r| )}{\gamma |r|}} &\;\;\;\; \mbox{d=3},
\end{eqnarray}
%
%
where $K_0$ is the modified Bessel function.
The constant $\gamma$ gives the inverse of the coupling length, and is
calculated as
\begin{equation}
  \gamma = \sqrt{\frac{\eta}{D}}.
\end{equation}
Each $G(r)$ must satisfy the normalization condition $\int G(r) d^d
{\bi r} = 1$.
Since we treat a two-dimensional system, we use Eq.(\ref{Eq:Kernel2d})
for $G(r)$ hereafter.

As elements, we use complex Ginzburg-Landau oscillators. They are the
simplest limit cycle oscillators that can be derived by the
center-manifold reduction technique from generic oscillators near
their Hopf bifurcation points\cite{Kuramoto2}.
The corresponding non-locally coupled system is given by the following
equation for the complex amplitude $W$:
\begin{equation}
  \dot{W}({\bi r}, t) = W - (1+ic_2)|W|^2 W + K (1+ic_1) (\bar{W} - W),
\end{equation}
where $K$ is the coupling strength, $c_1, c_2$ are real parameters,
and the non-local mean field $\bar{W}$ is given by
\begin{equation}
  \bar{W}({\bi r}, t) = \int d{\bi r'} G({\bi r'} - {\bi r}) W({\bi r'}, t).
\end{equation}
This is the non-local complex Ginzburg-Landau equation introduced by
Kuramoto\cite{Kuramoto1} as the first concrete example of non-locally
coupled systems.


\section{Anomalous spatio-temporal chaos}

In the numerical simulations presented here, we assume the total
system to be a square lattice with both sides unit length long.
The elements are placed on the lattice sites, and periodic boundary
conditions are assumed.
We use $N^2=512^2 \sim 1024^2$ elements, and fix the coupling length
$\gamma^{-1}$ at $1/8$.
The non-local mean field is easily calculated by using the FFT
technique, since it is simply a convolution of the amplitude field
with the kernel~(\ref{Eq:Kernel2d}).
We fix the parameters $c_1$ and $c_2$ at $-2$ and $2$ respectively.
These are the standard values already used in our previous
one-dimensional simulations.

\begin{figure}[htbp]
  \begin{center}
    \leavevmode
    \epsfxsize=0.4\textwidth
    \epsfbox{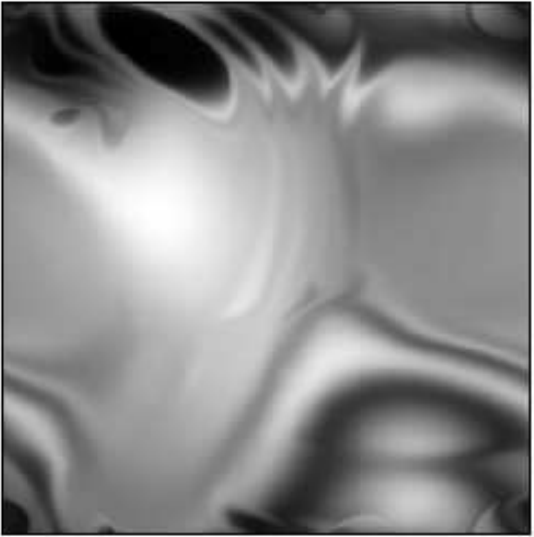}
    \caption{Snapshot of the amplitude field $X(x,y)$ at $K=1.05$.
      Darker dots indicates larger amplitudes.}
    \label{Fig:01}
  \end{center}
\end{figure}

\begin{figure}[htbp]
  \begin{center}
    \leavevmode
    \epsfxsize=0.4\textwidth
    \epsfbox{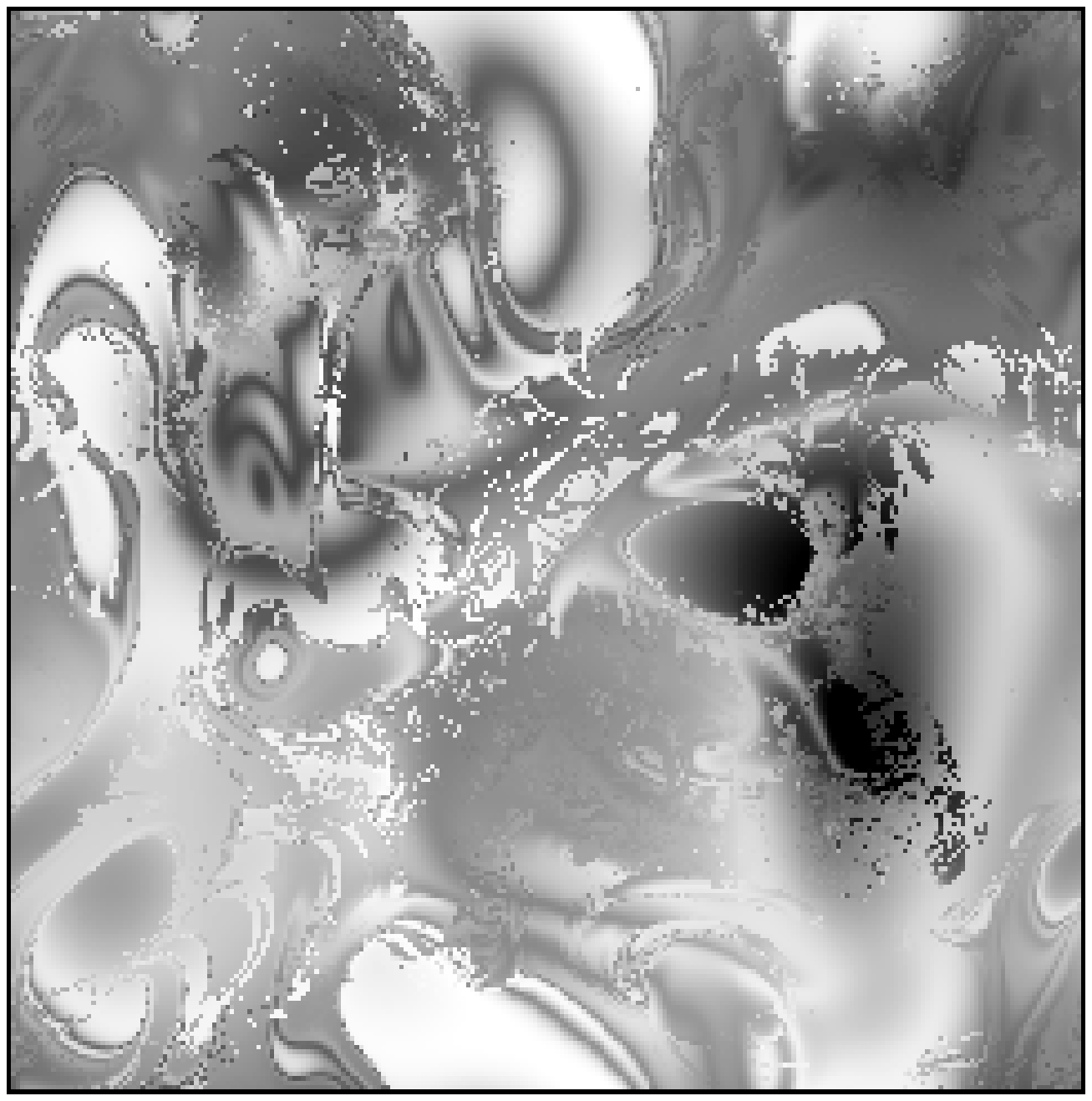}
    \caption{Snapshot of the amplitude field $X(x,y)$ at $K=0.85$.}
    \label{Fig:02}
  \end{center}
\end{figure}

\begin{figure}[htbp]
  \begin{center}
    \leavevmode
    \epsfxsize=0.4\textwidth
    \epsfbox{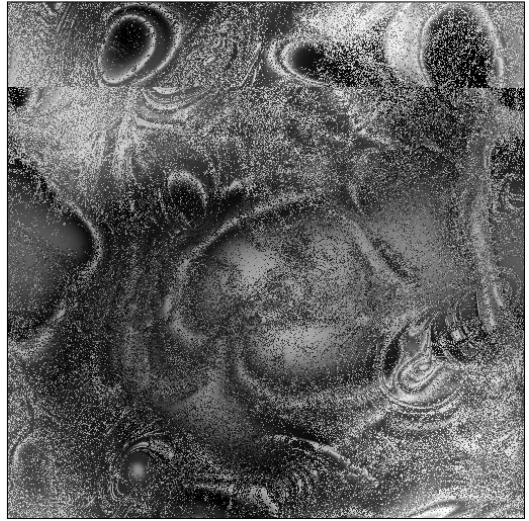}
    \caption{Snapshot of the amplitude field $X(x,y)$ at $K=0.65$.}
    \label{Fig:03}
  \end{center}
\end{figure}

In Fig.\ref{Fig:01}-\ref{Fig:03}, typical snapshots of the real part
$X(x, y)$ of the complex variable $W(x, y)$ are shown for three
different values of coupling strength $K$.
Since we obtain similar figures for the imaginary part $Y(x,y)$ by
symmetry, we use $X(x,y)$ in the following analysis and call it
amplitude field.
The amplitude field at $K=1.05$ is continuous and smooth, while at
$K=0.65$ it seems to be discontinuous and disordered, although not
completely random.
The amplitude field at the intermediate coupling strength $K=0.85$
looks somewhat more complex and intriguing; it is composed of
intricately convoluted smooth and disordered patches of various length
scales.
This is the anomalous spatio-temporal chaotic regime that our interest
is focused on.

\vspace{1.0cm}

\section{Spatial correlation function}

\begin{figure}[htbp]
  \begin{center}
    \leavevmode
    \epsfxsize=0.4\textwidth
    \epsfbox{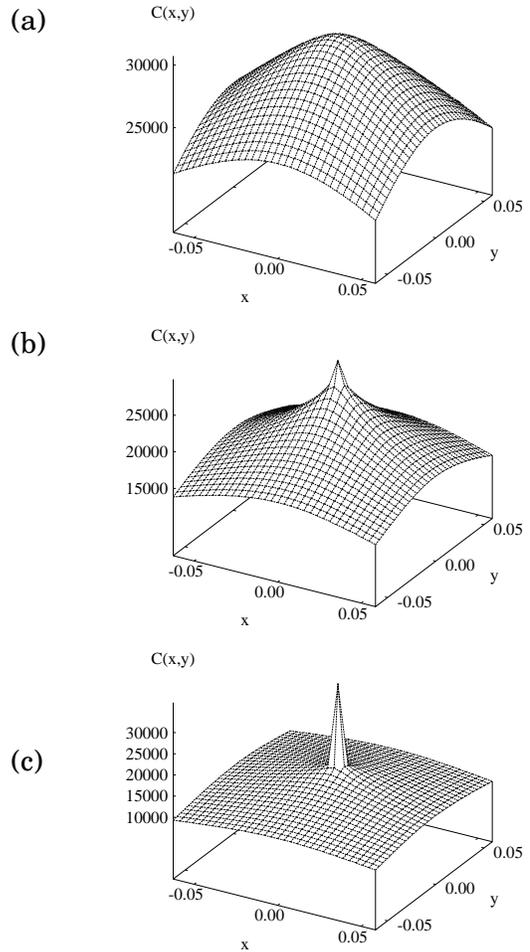}
    \caption{Short-range spatial correlation functions $C(x,y)$
      at (a) $K=1.05$, (b) $K=0.85$, and (c) $K=0.65$.}
    \label{Fig:04}
  \end{center}
\end{figure}

Let us examine the spatial correlation function first.
Figures~\ref{Fig:04}(a)-(c) show spatial correlation functions $C(x,y)
:= \langle X(0,0) X(x,y) \rangle$ corresponding to the amplitude
fields shown in Figs.\ref{Fig:01}-\ref{Fig:03}.
Each correlation function is clearly rotationally symmetric due to the
isotropy of the system.
As the amplitude field becomes disordered, the correlation function
becomes steep, and the center of the graph, which corresponds to the
self-correlation $C(0,0)$, becomes peaked.

In Ref.\cite{Kuramoto1}, the anomalous spatio-temporal chaotic regime
was characterized by the power-law behavior of the spatial correlation
function in small distance:
\begin{equation}
  C(l) := \langle X(0) X(l) \rangle \simeq C_0 - C_1 l^{\alpha}
  \;\;\;\; (l \ll 1),
\end{equation}
where $C_0, C_1$ are constants and $\alpha$ is a non-integer
parameter-dependent exponent.

\begin{figure}[htbp]
  \begin{center}
    \leavevmode
    \epsfxsize=0.4\textwidth
    \epsfbox{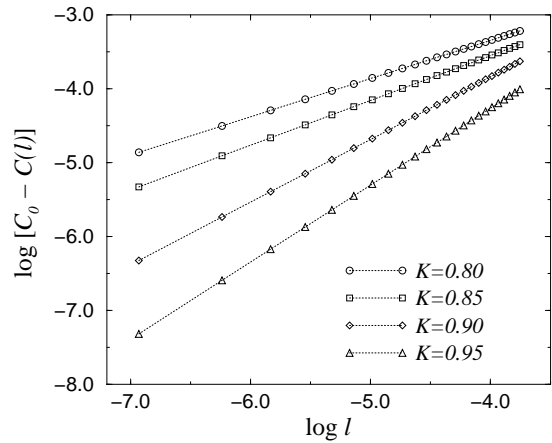}
    \caption{Power-law behavior of the correlation functions;
      log-log plot of $C_0-C(l)$ vs. $l$ obtained for several values
      of the coupling strength $K$. }
    \label{Fig:05}
  \end{center}
\end{figure}

To confirm if this power-law behavior also holds in two dimension, we
calculated the radial correlation function $C(l) = \langle X({\bi r})
X({\bi r}+{\bi l}) \rangle$ ($|{\bi l}| = l$) along a straight line in
a certain direction\footnote{We mainly used the $(0,1)$ or the $(1,1)$
  direction, but results are independent of the direction. }, and
estimated the best fitting parameters $C_0$ and $C_1$.
Figure~\ref{Fig:05} shows $\ln l$ vs. $\ln \left[ C_0 - C(l) \right]$
for several values of the coupling strength $K$.
For each coupling strength, the experimental data are almost on a
straight line, and the power-law behavior is evident.
The exponent $\alpha$ of the power law varies continuously with the
coupling strength.
Although not shown in the figure, the correlation function $C(l)$ is
continuous at the origin $l=0$ for $K \ge 0.85$, but discontinuous at
$K=0.80$. There appears a finite gap between the self-correlation
$C(0)$ and the correlation between the nearest-neighbor elements
$\lim_{l \to +0} C(l) \simeq C_0$. This means that the individual
motion of the element becomes so violent that the amplitude field is
no longer continuous statistically.
In Ref.\cite{Nakao1}, this transition point was identified with the
blowout bifurcation point in the on-off intermittent dynamics of the
amplitude difference between nearby elements.

Thus, the anomalous spatio-temporal chaos in two dimension is also
characterized by power-law behavior of the spatial correlation
function with a parameter-dependent exponent.

\section{Noisy on-off intermittency}

In Ref.\cite{Kuramoto3,Nakao1}, we clarified that the scaling behavior
of the spatial correlation is a consequence of underlying
multiplicative processes of amplitude differences between neighboring
elements.
We described this process by a multiplicative stochastic model, and
related the exponent $\alpha$ of the spatial correlation with the
fluctuation of the finite-time Lyapunov exponent of the element.
Such a model is essentially identical with those used in describing
noisy on-off intermittency\cite{Pikovsky,Platt,Cenys}, which indicates
that our system also exhibits this type of temporal intermittency.
Actually, the finite-time Lyapunov exponent of the complex
Ginzburg-Landau oscillator can fluctuate between positive and negative
values, and neighboring oscillators are subjected to only slightly
different non-local mean field.
Therefore, the conditions for the appearance of noisy on-off
intermittency are satisfied in our system.

\begin{figure}[htbp]
  \begin{center}
    \leavevmode
    \epsfxsize=0.4\textwidth
    \epsfbox{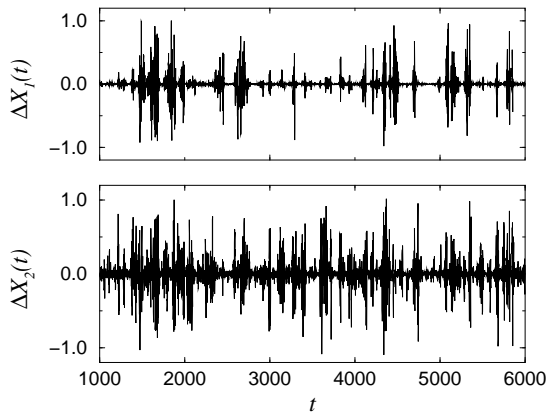}
    \caption{Typical temporal evolution of the amplitude differences.
      The distance between the elements is $512^{-1}$ for $\Delta
      X_1(t)$, and $64^{-1}$ for $\Delta X_2(t)$.}
    \label{Fig:06}
  \end{center}
\end{figure}

Now, let us confirm this in our system numerically. The coupling
strength is set at $K=0.85$.
Figure~\ref{Fig:06} shows typical time sequences of amplitude
differences $\Delta X_1(t)$ and $\Delta X_2(t)$. The distance between
the elements is $512^{-1}$ for $\Delta X_1(t)$, and $64^{-1}$ for
$\Delta X_2(t)$, respectively.
Strong intermittency of the signals is apparent. It can be seen that
$\Delta X_2(t)$ shows more frequent bursts than $\Delta X_1(t)$,
reflecting that $\Delta X_2(t)$ is subjected to larger fluctuations
than $\Delta X_1(t)$.

\begin{figure}[htbp]
  \begin{center}
    \leavevmode
    \epsfxsize=0.4\textwidth
    \epsfbox{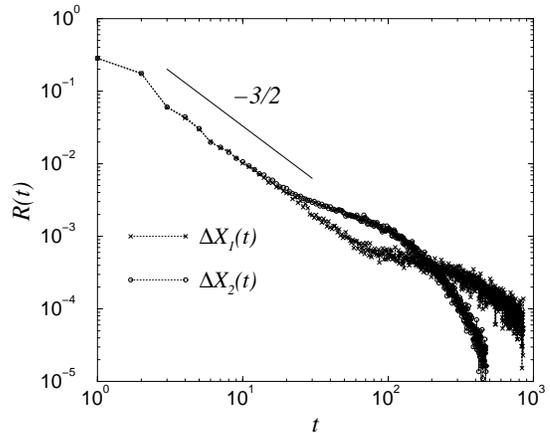}
    \caption{Distributions of the laminar length obtained from the time sequences
      $\Delta X_1(t)$ and $\Delta X_2(t)$ shown in Fig.\ref{Fig:06}.}
    \label{Fig:07}
  \end{center}
\end{figure}

We can confirm that these intermittent signals are actually noisy
on-off intermittent by calculating the laminar length distribution.
The laminar phase is defined as a successive duration where the
absolute value of the difference does not exceeds a certain threshold.
Here we choose $0.5$ as the threshold value.
Figure~\ref{Fig:07} shows laminar length distributions $R(t)$ obtained
from $\Delta X_1(t)$ and $\Delta X_2(t)$.
The characteristic shape of the distribution $R(t)$, i.e., the
power-law dependence on $t$ with slope $-3/2$ for small $t$, and the
exponential shoulder seen in the large $t$ region, clearly indicates
that the signals are actually noisy on-off intermittent.
The shoulder reflects broken scale invariance due to the additive
noise. As expected, the shoulder of $\Delta X_2(t)$ appears at a
smaller value of $t$ than that of $\Delta X_1(t)$\cite{Platt,Cenys}.

\section{Multi-scaling analysis}

The notion of multi-scaling, i.e., multi-affinity and
multi-fractality, have been employed successfully in characterizing
complex spatio-temporal behavior of various phenomena, such as
velocity and energy dissipation fields in fluid
turbulence\cite{Bohr,Frisch}, rough interfaces in fractal surface
growth\cite{Meakin}, nematic fluid electro-convective
turbulence\cite{Carbone}, financial data of currency exchange
rates\cite{Vandewalle}, and even in natural images\cite{Turiel}.
In Ref.\cite{Kuramoto3,Nakao2}, we introduced multi-scaling analysis
to our system for the one-dimensional case, inspired by the seeming
similarity of the amplitude and difference fields in our system to the
velocity and energy dissipation fields in fluid turbulence.
Here, we attempt the multi-scaling analysis for the two-dimensional
case.

\begin{figure}[htbp]
  \begin{center}
    \leavevmode
    \epsfxsize=0.4\textwidth
    \epsfbox{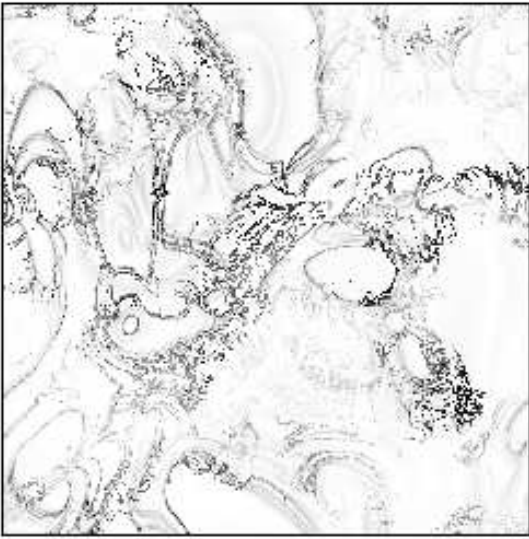}
    \caption{Snapshot of the difference field $Z(x,y)$ at $K=0.85$,
      corresponding to the amplitude field shown in Fig.\ref{Fig:02}.}
    \label{Fig:08}
  \end{center}
\end{figure}

First, we introduce the difference field $Z({\bi r})$ as
\begin{equation}
  Z({\bi r}) := | \nabla X({\bi r}) | = \sqrt{ \left( \frac{\partial
        X}{\partial x} \right)^2 + \left( \frac{\partial X}{\partial
        y} \right)^2},
\end{equation}
which emphasizes the edges of the original amplitude field $X({\bi
  r})$ \footnote{Here, the differential should not be interpreted
  literally. We always use finite difference in the actual
  calculation, e.g. $(X(x+\Delta x, y) - X(x, y))/\Delta x$ with
  sufficiently small $\Delta x$, and this is important to observe the
  multi-scaling behavior\cite{Nakao2}.}.  This quantity is an analogue
of the energy dissipation field in fluid turbulence.
Figure~\ref{Fig:08} shows a typical snapshot of the difference field
$Z(x,y)$ at $K=0.85$, corresponding to the amplitude field shown in
Fig.\ref{Fig:02}. The intermittency underlying the original amplitude
field is now apparent.

We then introduce the following quantities as measures for the
amplitude field $X({\bi r})$ and the difference field $Z({\bi r})$:
\begin{eqnarray}
  h({\bi r} ; l) &:=& | X({\bi r}+{\bi l}) - X({\bi r}) |, \\
  m({\bi r} ; l) &:=& \int_{S({\bi r} ; l)} Z({\bi r}') d^2{\bi r}',\label{Eq:ml}
\end{eqnarray}
where $|{\bi l}| = l$, and the domain of integration $S({\bi r} ; l)$
is a square of size $l$ placed at ${\bi r}$.
The first quantity is a difference of the amplitude field $X({\bi r})$
between two points separated by a distance of $l$, and the second
quantity is a volume enclosed by the difference field $Z({\bi r})$ and
the square $S({\bi r}; l)$.

According to the multi-fractal formalism, two types of partition
functions are defined as
\begin{eqnarray}
  Z_h^q(l) &:=& \langle h(l)^q \rangle = \frac{1}{M(l)}
  \sum_{i=1}^{M(l)} h({\bi r}_i ; l)^q, \\ 
  Z_m^q(l) &:=& N(l) \langle m(l)^q \rangle = \sum_{i=1}^{N(l)} m({\bi
    r}_i ; l)^q,
\end{eqnarray}
where $Z_h^q(l)$ is calculated along a certain straight line in some
direction as in the case of the previous spatial correlation function,
while $Z_m^q(l)$ is calculated over the whole system. ${\bi r}_i$ is
either the position of the line segment or the position of the square.
$M(l)$ is the number of line segments of length $l$ that are needed to
cover the entire line, and $N(l)$ is the number of squares of size $l$
that are needed to cover the whole system. The function $Z_h^q(l)$ is
called structure function in the context of fluid
turbulence\cite{Bohr,Frisch}.

When the measures have scaling properties, the partition functions are
expected to scale with $l$ as $Z_h^q(l) \sim l^{\zeta(q)}$ and
$Z_m^q(l) \sim l^{\tau(q)}$. Furthermore, if these exponents
$\zeta(q)$ and $\tau(q)$ depend nonlinearly on $q$, the corresponding
measures $h(l)$ and $m(l)$ are called multi-affine and multi-fractal,
respectively\cite{Bohr,Frisch,Meakin}.

For the one-dimensional case, we already know that the amplitude field
is multi-affine and the difference field is multi-fractal.
Moreover, our previous theory predicts the following form for the
scaling exponent $\zeta(q)$ of the amplitude field:
\begin{equation}
  \zeta(q) = q \; (0 < q < \beta), \;\;\;\; \beta \; (\beta < q),
  \label{Eq:bifractal}
\end{equation}
%
%
where $\beta$ is a positive constant determined by the fluctuation of
the finite-time Lyapunov exponent of the element, and is related to
the slope of the probability distribution of
$h(l)$\cite{Kuramoto3,Nakao2}.
This is the simplest form of multi-affinity, and sometimes called
bi-fractality\cite{Frisch}\footnote{There is some confusion in
  terminology due to historical reasons. A word 'bi-affinity' will be
  more appropriate if it exists.}.
The same form of the scaling exponent $\zeta(q)$ is also expected in
two dimension, since our previous theory imposed no restriction on the
dimensionality of the system.

For the scaling exponent $\tau(q)$ of the difference field, we have
not been able to develop a satisfactory theory yet. Numerical results
in one-dimensional systems suggest that $\tau(q)$ also depends
nonlinearly on $q$, and the difference field is multi-fractal with a
rather simple functional form for $\tau(q)$\cite{Nakao2}.
However, the scaling exponent $\tau(q)$ for the two-dimensional system
may be different from that for the one-dimensional case, since
$Z_m^q(l)$ is defined depending on the dimensionality of the system,
while $Z_h^q(l)$ is always defined along a one-dimensional line.

\begin{figure}[htbp]
  \begin{center}
    \leavevmode
    \epsfxsize=0.4\textwidth
    \epsfbox{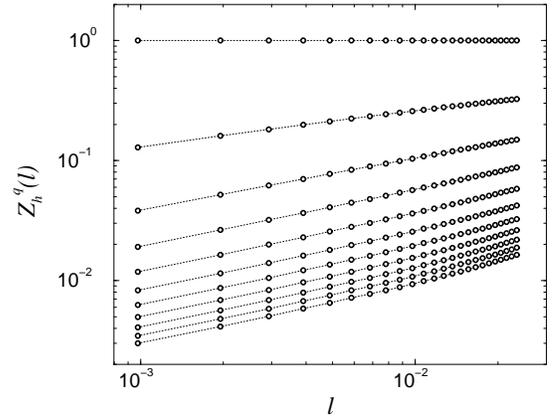}
    \caption{Partition functions $Z_h^q(l)$ vs. $l$ for several values of $q$.
      The top line corresponds to $q=0$, and the bottom one to $q=5$,
      at intervals of 0.5.}
    \label{Fig:09}
  \end{center}
\end{figure}

\begin{figure}[htbp]
  \begin{center}
    \leavevmode
    \epsfxsize=0.4\textwidth
    \epsfbox{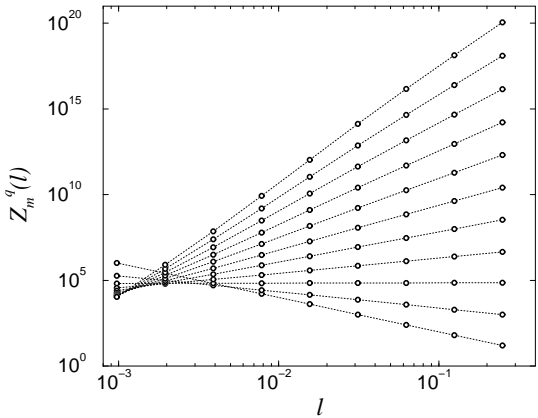}
    \caption{Partition functions $Z_m^q(l)$ vs. $l$ for several values of $q$.
      The bottom line corresponds to $q=0$, and the top one to $q=5$,
      at intervals of 0.5.}
    \label{Fig:10}
  \end{center}
\end{figure}

Let us proceed to the numerical results now. The coupling strength is
fixed at $K=0.85$ hereafter, where the system is fully in the
anomalous spatio-temporal chaotic regime.

\begin{figure}[htbp]
  \begin{center}
    \leavevmode
    \epsfxsize=0.4\textwidth
    \epsfbox{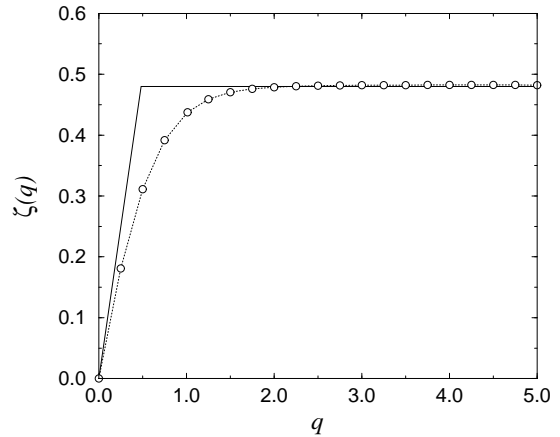}
    \caption{Scaling exponent $\zeta(q)$ vs. $q$.
      The theoretical curve Eq.(\ref{Eq:bifractal}) with $\beta=0.48$
      is compared with the experimental data.}
    \label{Fig:11}
  \end{center}
\end{figure}

Figure~\ref{Fig:09} shows the partition function $Z_h^q(l)$ obtained
for several values of $q$. Each curve depends on $l$ in a power-law
manner for small $l$, and its exponent increases with $q$.
The dependence of the scaling exponent $\zeta(q)$ on $q$ is shown in
Fig.\ref{Fig:11}. The $\zeta(q)$ curve is a strongly nonlinear
function of $q$, and the multi-scaling property of the amplitude field
is evident.
Furthermore, the $\zeta(q)$ curve has a bi-linear form as expected in
Eq.(\ref{Eq:bifractal}), although a sharp transition is absent due to
the limited number of oscillators.
From the large $q$ behavior of the exponent, we can roughly estimate
the value of $\beta$ as $\sim 0.48$.

Thus, the amplitude field turns out to be multi-affine, and the
behavior of the scaling exponent is the same as that for the
one-dimensional case.

\begin{figure}[htbp]
  \begin{center}
    \leavevmode
    \epsfxsize=0.4\textwidth
    \epsfbox{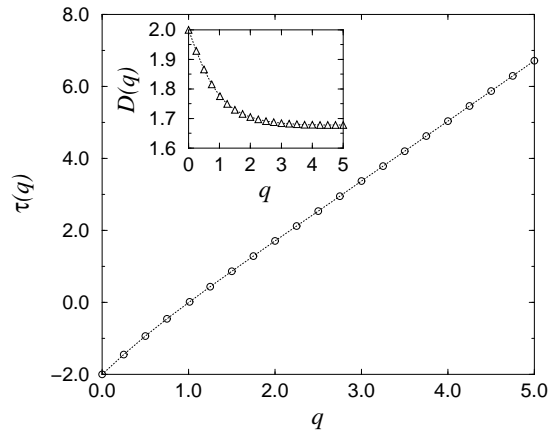}
    \caption{Scaling exponent $\tau(q)$ vs. $q$.
      The inset shows the corresponding generalized dimension $D(q)$
      vs. $q$.}
    \label{Fig:12}
  \end{center}
\end{figure}

Figure~\ref{Fig:10} shows the partition function $Z_m^q(l)$ for
several values of $q$. It is clear that each curve shows a power-law
dependence on $l$. The width of the region where the power law holds
seems much wider than the previous case.
The scaling exponent $\tau(q)$ is plotted with regard to $q$ in
Fig.\ref{Fig:12}. The corresponding generalized dimension $D(q) :=
\tau(q) / (q-1)$ is also shown in the inset.
The $\tau(q)$ curve is again a nonlinear function of $q$, but its
dependence on $q$ does not seem to be so simple as that of the
$\zeta(q)$ curve.
However, as we conjectured numerically for the one-dimensional
case\cite{Nakao2}, asymptotic linearity of the $\tau(q)$ curve seems
to hold.
Correspondingly, the $D(q)$ curve seems to saturate to a horizontal
line $D(q) = D(\infty)$ quickly.
But we can not observe a clear transition to the horizontal line as in
the previous one-dimensional case, which may be due to the limited
number of oscillators, or the two-dimensionality of the system.

Thus, the difference field also turns out to be multi-fractal. The
behavior of the scaling exponent somewhat resembles that in the
one-dimensional case, but is not completely the same.

\section{Probability distributions of the measures}

The multi-scaling properties of amplitude and difference fields are
consequences of intermittency underlying the system.
In order to analyze this intermittency in more detail, we study here
the probability distribution functions (PDF) of both measures at each
length scale.

Let us consider the PDFs of the measures $h(l)$ and $m(l)$.
It is convenient to use rescaled measures $h_r(l) : = h(l) / l$,
$m_r(l) := m(l) / l^2$ and corresponding rescaled PDFs $P_r(h_r ; l)$,
$Q_r(m_r ; l)$.
With this rescaling, the peaks and the widths of the PDFs become
relatively close. We show our results in these rescaled variables.

The scaling exponents $\zeta(q)$ and $\tau(q)$ are fully determined by
the dependence of these PDFs on the scale $l$.
In Ref.~\cite{Nakao2}, we approximated the tails of the PDFs by
certain functional forms and extracted the scaling exponents
asymptotically in the small $l$ limit.
Here we present the numerical results only briefly for the purpose of
emphasizing the intermittency of the amplitude and difference fields.

\begin{figure}[htbp]
  \begin{center}
    \leavevmode
    \epsfxsize=0.4\textwidth
    \epsfbox{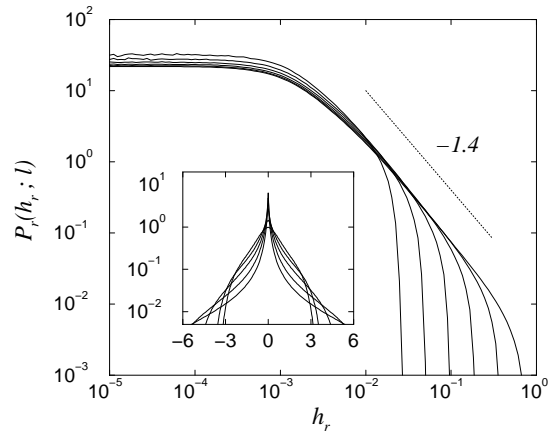}
    \caption{Rescaled PDFs $P_r(h_r ; l)$ for $l=\sqrt{2}$, $2\sqrt{2}$,
      $4\sqrt{2}$, $8\sqrt{2}$, $16\sqrt{2}$, and $32\sqrt{2}$
      ($\times 1024^{-1}$). The curve with the slowest cutoff
      corresponds to $l=\sqrt{2} \times 1024^{-1}$, and the leftmost
      curve with the fastest cutoff corresponds to $l=32\sqrt{2}
      \times 1024^{-1}$. The inset shows PDFs of the amplitude
      difference $h(l)$ (without taking the absolute value) rescaled
      by the standard deviation in log-linear scales for $l=\sqrt{2}$
      (the steepest curve), $4\sqrt{2}$, $16\sqrt{2}$, $64\sqrt{2}$,
      and $128\sqrt{2}$ (the nearly quadratic curve) ($\times
      1024^{-1}$).}
    \label{Fig:13}
  \end{center}
\end{figure}

Figure~\ref{Fig:13} shows the rescaled PDF $P_r(h_r ; l)$ of the
rescaled amplitude difference $h_r(l)$ for several values of $l$ in
log-log scales.
Each PDF has a characteristic truncated L\'evy-like shape, as we
already obtained previously for the one-dimensional case; it is
composed of a constant region near the origin, a power-law decay in
the middle, and a sharp cutoff.
Each curve roughly collapses to a scale-invariant curve in the
constant and power-law regions, while the cut-off of the tail moves to
the right with decreasing of $l$ due to the intermittency.
More precisely, the cut-off position of the PDF (defined in some
suitable way) is proportional to $l^{-1}$. This gives rise to the
bi-fractal behavior of the $\zeta(q)$ curve, see Ref.~\cite{Nakao2}
for detail.

Our previous theory predicts that the slope of the power-law decay is
given by $-1-\beta$ with the constant $\beta$ from
Eq.~(\ref{Eq:bifractal}). This is confirmed in Fig.\ref{Fig:13}, where
the slope of the power-law decay can be read off as $-1.4$, roughly in
agreement with the previously obtained value $\beta \sim 0.48$ from
the scaling exponent $\zeta(q)$.
The inset of Fig.~\ref{Fig:13} shows the PDF of the amplitude
difference $h(l)$ (without taking the absolute value) rescaled by the
standard deviation in linear-log scales, in order to further emphasize
the intermittency of the amplitude field. The PDF evolves from nearly
Gaussian into intermittent power-law with the decrease of scale $l$.

\begin{figure}[htbp]
  \begin{center}
    \leavevmode
    \epsfxsize=0.4\textwidth
    \epsfbox{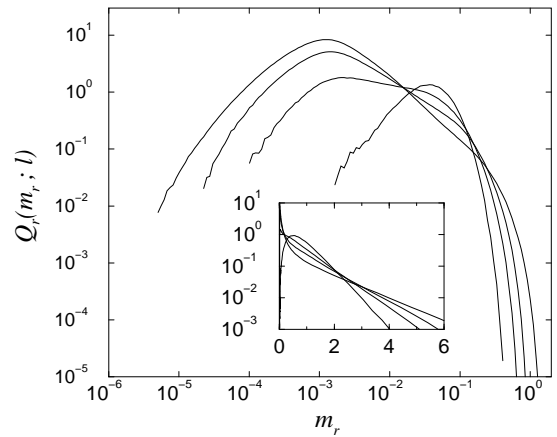}
    \caption{Rescaled PDFs $Q_r(m_r ; l)$ for $l=2$
      (the widest distribution with the steepest power-law decay),
      $8$, $32$, $128$ (the narrowest) ($\times 1024^{-1}$). The inset
      shows PDFs of $m(l)$ rescaled by the standard deviation in
      log-linear scales for the same set of $l$ values. The steepest
      curve corresponds to $l=2 \times 1024^{-1}$, and the curve with
      nearly quadratic peak corresponds to $l=128 \times 1024^{-1}$.}
    \label{Fig:14}
  \end{center}
\end{figure}

Figure~\ref{Fig:14} shows the PDFs $Q_r(m_r ; l)$ of the rescaled
volume $m_r(l)$. Their shapes are not so simple as those for $P_r(h_r ;
l)$.
They are also qualitatively different from the PDF for the difference
field in the previous one-dimensional case\cite{Nakao2}, which is
responsible for the slightly different behavior of the scaling
exponent $\tau(q)$ from the one-dimensional case.
But we can still see that both tails extend, and the distribution
widens with the decrease of scale $l$. Namely, as we decrease the
observation scale $l$, largely deviated events appear more frequently.
It is obvious that this intermittency effect gives rise to the
nonlinearity of the scaling exponent $\tau(q)$, although the precise
functional form is difficult to obtain.
The inset shows the PDF of the measure $m(l)$ rescaled by the standard
deviation in linear-log scales for several values of $l$. The PDF
gradually gets steeper with the decrease of $l$ due to the
intermittency.

Thus, the PDFs of the measures reveal the intermittency in our system
clearly. Especially, the PDF for the amplitude difference $h(l)$ has
the same shape as that already obtained in the one-dimensional case.

\section{Conclusion}

We numerically analyzed a two-dimensional system of non-locally
coupled complex Ginzburg-Landau oscillators.
As in the previous one-dimensional case, we found an anomalous
spatio-temporally chaotic regime characterized by a power-law behavior
of the spatial correlation function.
As expected from our previous theory, the amplitude difference between
neighboring elements exhibits noisy on-off intermittency, giving a
microscopic dynamical origin for the power-law spatial correlations.
We performed multi-scaling analysis in that regime, and found that the
amplitude and difference fields are indeed multi-affine and
multi-fractal, indicating strong intermittency underlying the system.
By studying the PDFs of the measures at each length scale, the
intermittency was clearly observed as scale-dependent deviations of
the PDFs in their tails.

Multi-scaling properties are also known in various phenomena such as
turbulence or fractal surface growth. The appearance of similar
multi-scaling properties in many different systems implies some
underlying common statistical law leading to such behavior.
Further study of the intermittency in our system will give more
insights and hints for the understanding of the multi-scaling
properties observed in complex dissipative systems.

\acknowledgements

The author gratefully acknowledges helpful advice and continuous
support by Prof. Yoshiki Kuramoto, and thanks Dr. Axel Rossberg for a
critical reading of the manuscript.
He also thanks the Yukawa Institute for providing computer resources,
and the JSPS Research Fellowships for Young Scientists for financial
support.

\end{document}